\documentclass{elsart}
\usepackage{epsfig}
\usepackage{rotate}
\usepackage{amssymb}
\usepackage{amsmath}

\usepackage{bm}% bold math
\usepackage{amssymb}
\usepackage{enumerate}

\newcommand\klinv{K_L \to invisible}

\begin{document}
\begin{frontmatter}
\title{\boldmath Invisible  $K_L$ decays in the SM extensions}
\begin{center}
S.N.~Gninenko$^{1}$ and  N.V.~Krasnikov$^{1,2}$ 
\vskip0.5cm
$^{1}${\em Institute for Nuclear Research, Moscow 117312}\\
$^{2}$ {\em Joint Institute for Nuclear Research, 141980 Dubna, Russia}
\end{center}
\begin{abstract}
In the Standard Model (SM) the  branching ratio for the decay   $K_L \to \nu \bar{\nu}$  
into two neutrinos is helicity suppressed and predicted to be very small  
$  \leq O(10^{-17})$. 
We consider two natural extensions  of the SM, such as two-Higgs-doublet model (2HDM) and 
the $\nu$MSM with additional singlet scalar, those main 
features is that they  can lead to an enhanced
$Br(\klinv)$. In the  2HDM  the smallness of the neutrino mass is explained 
due to the smallness of the second Higgs doublet vacuum expectation value.   
 Moreover, the $\nu$MSM extension with 
additional singlet field can explain the  $(g - 2)$ muon anomaly. 
The considered models 
demonstrate  that the $\klinv$ decay is a clean probe of new physics scales well above 100 TeV, that is complementary to rare 
$K \to \pi+invisible$ decay,  and provide a strong  motivation for its sensitive search in a near future low-energy experiment. 
\end{abstract}
\end{frontmatter}

%\pacs{14.80.-j, 12.60.-i, 13.20.Cz, 13.35.Hb}% PACS, the Physics and Astronomy
%\maketitle

\section{Introduction}

The discovery of the neutrino oscillations \cite{Neutrino,pdg} means that at least two neutrino have nonzero masses. 
The minimal extension of the 
SM with nonzero neutrino masses  is the $\nu$MSM model \cite{nuMSM,nuMSM1}. In this model one adds to the SM 
three additional massive Majorana(sterile) fermions $\nu_{Ri}$, $i = 1,2,3$. 
Due to the seesaw mechanism \cite{nuMSM,seesaw} after the spontaneous 
$SU_L(2) \otimes U(1)$ electroweak symmetry breaking the neutrinos acquire masses $m_{\nu_i}= \frac{m^2_{Di}}{M_{Ri}}$. 
Here $m_{Di}$ are the Dirac neutrino masses and  $M_{Ri}$ are the Majorana masses of the sterile $\nu_{Ri}$ neutrinos. 
The $\nu$MSM with relatively light neutrino Majorana masses  
 $M_{\nu_{Ri}} \leq O(5)~GeV$ has a candidate - 
the lightest neutrino Majorana with a mass  $M_{\nu_{R}} \leq O(50)~KeV$ - for  dark matter. 
Besides, the model 
with light Majorana neutrino can solve the problem of the baryon asymmetry in our Universe \cite{nuMSM1}. 

If masses of Majorana neutrinos are lying in sub-GeV region,  the decays of neutral mesons into neutrino pairs, such e.g. as $\pi^0, \eta, \eta', K_S, K_L\to \nu \overline{\nu}$ decays can occur. 
Since the neutrinos are weekly interacting particles 
and do not  interact in the detector, such  decays are invisible, thus making  their searches  extremely difficult task.
In particular, the decay $\klinv$ has never been experimentally tested.
The branching ratio of the $\klinv$ decay  in the SM is predicted to be very small 
$\leq O(10^{-17})$
 for $\nu$ masses laying in the  sub-eV region favored by the observations of
$\nu$ oscillations \cite{Neutrino,pdg}.   
Indeed, the $K_L$ has zero spin, and it cannot decay into two massless neutrinos, 
as it contradicts to momentum and angular momentum conservation simultaneously. 
Therefore, an observed $Br(\klinv) \geq 10^{-10}$ would unambiguously signal the presence of the BSM physics. 
Recently, an approach for performing such kind of experiments by using  
the $K^+ n \to  K^0 p$ (or  $K^-  p \to  \overline{K}^0 n$ ) charge-exchange reaction 
as a source of well tagged $K^0$'s has been reported and the first experimental bound $ Br(K_L \to invisible ) \lesssim 6.3 \cdot 10^{-4}$ 
has been set from existing experimental data \cite{Gninenko}. It has been shown, that 
compared to this limit,  the expected sensitivity of the proposed search is at least two orders of magnitude 
higher - $Br(K_L \to invisible) \lesssim 10^{-6}$  per $\simeq 10^{12}$ incident kaons. This  limit can be further  
improved by utilizing a  more detailed design of the experiment, thus making the region  
$Br(K_L \to invisible) \simeq 10^{-8} - 10^{-6}$, and  even below,  experimentally accessible \cite{Gninenko}. 
In Ref.\cite{GniKr} we considered several natural extensions  of the SM, such as two-Higgs-doublet (2HDM), 2HDM and light scalar, 
and mirror dark matter models, those main 
feature is that they allow to avoid the helicity suppression factor for the previously mentioned pseudoscalar mesons decays into neutrino  
and lead to an enhanced
$Br(\klinv)$.

In this note,  which is a continuation of the work of  Ref. \cite{GniKr},  we consider two   SM extensions 
that can lead to 
  invisible $K_L$ decays at an experimentally interesting level $Br(K_L \rightarrow invisible) \geq O(10^{-8})$. Namely, 
we consider  the 2HDM  and the $\nu$MSM   with additional scalar isosinglet field, those main  feature is that they  
can lead to an enhanced  $Br(\klinv)$. In the  2HDM  the smallness of the neutrino mass is explained 
due to the smallness of the second Higgs doublet vacuum expectation value. 
The nonzero and very small value of the second Higgs doublet can arise as a consequence of nonzero quark condensate. Moreover, the $\nu$MSM extension with 
additional singlet field is able to explain the observed $(g - 2)$ muon anomaly. 
The considered models 
demonstrate  that the $\klinv$ decay is a clean probe of new physics scales well above 100 TeV, that is complementary 
to rare $K \to \pi+invisible$ decay,  and provide a strong  motivation 
for its sensitive search in a near future experiment.

%%%The end of the introduction

%The Lagrangian of the $\nu$MSM is 
%\begin{equation}
%L_{\nu SM} = L_{SM} + L_N \,,
%\end{equation}
%where 
%\begin{equation}
%L_N = \bar{N}_{i} i 
%\hat{\partial}N_i -
%f_{ji}\bar{L}_jHN_{i} - \frac{M_{N_{i}}}{2}\bar{N}^C_iN_{i} + H.c. \,
%\end{equation}
%and $L_{SM}$ is the SM Lagrangian.  Here $M_{N_i}$ are Majorana masses, $f_{ij}$ are Yukawa couplings and 
%$L_1 = (\nu_{eL}, e_L)$, 
%$L_2 = (\nu_{\mu L}, \mu_L)$, 
%$L_3 = (\nu_{\tau L}, \tau_L)$, $H = (H^0, H^{-})$\footnote{In the unitare gauge 
%$H = (\frac{v +h}{\sqrt{2}}, 0)$, $v = 246~GeV$}. 

%%% THE FIRST MODEL
%%%

\section{ $K_L \rightarrow \nu\bar{\nu}$ decay   in the two-Higgs-doublet model}

Consider  the $K_L \rightarrow \nu\bar{\nu}$ decay in {\it the two-Higgs-doublet model} (2HDM). 
The  2HDM  can have tree level  flavor-changing neutral currents, 
provide explanation of the origin of Dark Matter and $CP$ violation , see e.g. ref.\cite{branco}. 
The Lagrangian of our variant of the 2HDM has the form
 \begin{equation}
L_{tot} = L_{SM} + L_{H_2q} + L_{H_2\nu} +L_{HH_2}  \,,
\end{equation} 
where
\begin{equation}
L_{H_2q} =  -h_{Qd,ij}\bar{Q}_{Li}\bar{H}_2d_{Rj} +H.c. \,,
\end{equation}
\begin{equation}
L_{H_2 \nu} = i\bar{\nu_{Rj}} \hat{\partial} \nu_{Rj} -(\frac{M_{\nu, ij}}{2}\nu_{Ri}\nu_{Rj} 
+ h_{L\nu, ij}\bar{L}_iH_2\nu_{Rj}  + H.c.) \,,
\end{equation}
\begin{equation}
L_{HH_2} = \Delta^{\mu}H^*_2\Delta_{\mu}H_2 - M^2_{H_2}H^*_2H_2 -  \lambda_2(H^*_2H_2)^2    +  
(\delta m^2_{HH2}H^*H_2 +H.c.)\,
\end{equation}
and $L_{SM}$ is the SM Lagrangian. Here  $Q_{L1} = (u_L, d_L)$, $Q_{L2} = (c_L, s_L)$, 
$Q_{L3} = (t_L, b_L)$, $d_{R1} = d_R$, $d_{R2} = s_R$, $d_{R3} = b_R$, 
  $H_2 = (H^0_2, H^-_2)$,   $\bar{H}_2 = (-(H^-_2)^*, (H^0_2)^*)$, 
$\Delta_{\mu} = \partial_{\mu} + i\frac{g_1}{2}B_{\mu} - ig_2(\frac{\vec{\tau}}{2}\vec{Z}_{\mu})$. 

The neutrinos acquire  nonzero Dirac masses 
$m_{\nu, ij}^D = h_{L\nu,ij}<H_2>$
due to nonzero vacuum expectation value of the second Higgs isodoublet 
$<H_2> \approx  \frac{\delta m^2_{HH_2}}{M^2_{H_2}} <H> $  ($<H> = 174~GeV$). 
The smallness of the 
Dirac neutrino masses is a consequence of  the $<H_2>$  smallness. 
We can choose the  basis in which the Majorana mass matrix 
and the Yukawa coupling constants $  h_{L\nu, ij}$ are diagonal:
 $M_{\nu, ij} = M_{Ri}\delta^i_j$ ,  $ h_{L\nu, ij} =  h_{L,i}\delta^i_j$     and 
$M_{R1}$ has the minimal value. The value $M_{R1} = 0 $  corresponds to the case of Dirac neutrino.
Since we are interested mainly in the $K_L \rightarrow invisible$ decay we assume that  the decay 
$K_L \rightarrow \nu_{L1}\bar{\nu}_{R1}, \nu_{R1}\bar{\nu}_{L1} $ is kinematically allowed.

The effective four fermion Lagrangian describing the decay 
 $K_L \rightarrow \nu_{R1} \bar{\nu}_{L1}, \nu_{L1} \bar{\nu}_{R1} $  has the form 
\begin{equation}
L_{eff} = \frac{1}{M_{H_2}^2} [ h_{Qd,12}h_{L,1}\bar{d}_Ls_R\bar{\nu}_{L1}\nu_{R1} +
h_{Qd,21}^*h_{L,1}^*\bar{d}_Rs_L\bar{\nu}_{R1}\nu_{L1}]
+ H.c. \,.
\end{equation}
The decay rate     $K_L \rightarrow \nu_{R1} \bar{\nu}_{L1}, \nu_{L1} \bar{\nu}_{R1} $               is 
determined by the formula
\begin{eqnarray}
\Gamma(K_L \rightarrow \nu_{L1}\bar{\nu}_{R1}, \nu_{R1}\bar{\nu}_{L1}) =
\frac{M^5_{K_L}}{16\pi M^4_X} \nonumber \\
(\frac{F_K}{2(m_d + m_s)})^2 K(\frac{m^2_{R1}}{M^2_{K_L}}) \,,
\end{eqnarray}
where 
\begin{equation}
\frac{1}{M^4_X} = \frac{ |(h_{Qd,12} +  h_{Qd,21})h_{L,1 }|^2} {M^4_{H_2}}
\end{equation}
and   $K(x) = (1 - x)^2$   
for Majorana neutrino $\nu_{R1}$ with a mass $m_{R1}$ and massless $\nu_{L1}$ neutrino. 
Here  $F_K \approx 160~MeV$ is kaon decay constant  
and $m_s, m_d$ are the masses of $s$- and $d$-quarks\footnote{The quark masses $m_d,m_s$ and the effective 
mass $M_X$ implicitly depend on the renormalization point $\mu$ but  their combination $M^2_X(m_d + m_s)$ 
and hence the decay width (6)
is renormalization 
groop invariant and does not depend on the renormalization point $\mu$.}.   For 
$Br(K_L \rightarrow     \nu_{R1} \bar{\nu}_{L1}, \nu_{L1} \bar{\nu}_{R1}) = 10^{-6}$ we can test the value of $M_{X}$ up to
\footnote{In our estimate (8) we used 
the values $\tau(K_L) = 5.17\cdot 10^{-8}~sec$ and $(m_d + m_s) = 160~MeV$}
\begin{equation}
M_{X} \lesssim  0.6 \cdot 10^5~GeV
\end{equation}
for small neutrino mass $m_{R1} \ll M_{K_L}$. 

The Yukawa interaction (2)  leads to the effective 
tree level   flavour changing $\Delta S = 2$ effective interaction
\begin{equation}
L_{\Delta S = 2} =       \frac{1}{\Lambda^2_{\Delta S = 2}}\bar{d}_Ls_R\bar{d}_Rs_L + H.c.\,,
\end{equation}
where
\begin{equation}
\frac{1}{\Lambda^2_{\Delta S = 2}} =  \frac{h_{Qd,12}h^*_{Qd,21} }{M^2_{H2}} \,.
\end{equation}

The measured $K_L - K_S$ mass difference and the CP-violation parameter $\epsilon_K$ strongly restrict 
   \cite{Butler} the effective  $\Delta S = 2$ interaction (9), 
namely \cite{Butler}  
\begin{equation}
|Re(\Lambda_{\Delta S = 2})| \geq 1.8 \cdot 10^{7}~GeV \,,
\end{equation}
\begin{equation}
|Im(\Lambda_{\Delta S = 2})| \geq 3.2 \cdot 10^{8}~GeV \,.
\end{equation}
We shall consider the case of the CP-conserving interaction (2), i.e $h_{Qd,12}$ and $h_{Qd,21}$ are real. 
As a consequence of the inequality (11) and the formula (10) we find that
\begin{equation}
M_{H_2} \geq 1.8(|h_{Qd,21}h _{Qd,12}|)^{1/2} \cdot 10^{7}~GeV  \,.
\end{equation}
It should be noted that the bound (13) restricts rather strongly but not excludes 
the phenomenologically interesting values of the $M_X$.
Really, we  can simultaneously avoid the $\Delta S = 2$ bound (13)
  and obtain phenomenologically interesting values   
$Br(K_L \rightarrow \nu_{L1}\bar{\nu}_{R1}, \nu_{R1}\bar{\nu}_{L1}) $
for
small quark Yukawa coupling constants $h_{Qd,12},~h_{Qd, 21}$, relatively light  second Higgs doublet  
and not small lepton Yukawa coupling constant $h_{L,1}$ . For instance, for  $h_{Qd,12}   = h_{Qd,21}   = (1/300)^2$(
$h_{Qd,12} = 9h_{Qd,21} = 0.5 \cdot 10^{-4}$),
$h_{L,1} = 1$ and $M_{H_2} = 300~GeV$  we find that   $\Lambda_{\Delta s = 2} = 2.7 \cdot 10^{7}~GeV(1.8 \cdot 10^{7}~GeV) $
 and 
$Br(K_L \rightarrow \nu \bar{\nu}) =  0.4 \cdot10^{-6}(5 \cdot10^{-6}) $.

The existence of relatively light 
with a mass $M_{H_2} = 300~GeV$  second Higgs doublet does not contradict the LHC data. The best way to look for 
the second Higgs isodoublet at the LHC is the use of the reaction $pp \rightarrow Z^*/gamma^* \rightarrow H^+_2H^{-}_2 
\rightarrow l^+l^{-} \nu \bar{\nu}$ ($l^- = e, \mu, \tau $). So the signature is two  $l^+l^-$ leptons plus nonzero $E^T_{miss}$ in 
final state that coincides with the signature used for the search for direct production of sleptons at the LHC.

In considered model the neutrino $\nu_1$ acquires  nonzero Dirac mass 
$m_{\nu_{1}} = h_{L,1}<H_2>$
due to nonzero vacuum expectation value of the second Higgs isodoublet 
$<H_2> \approx  \frac{\delta m^2_{HH_2}}{M^2_{H_2}} <H> $  ($<H> = 174~GeV$). 
The smallness of the 
Dirac  neutrino mass is a consequence of  the $<H_2>$  smallness.
% and smallness of the Majorana mass $M_{R1}$. 
The smallness of $<H_2>$ is due to  the small value of 
$\delta m^2_{HH_2}$\footnote{In ref.\cite{Ma} a model with additional Higgs isodoublet interacting 
only with lepton fields was proposed. In this model the neutrinos acquire nonzero Dirac masses due to nonzero vacuum 
expectation value of the second Higgs isodoublet that allows to decrease the seesaw  scale from $O(10^{15})~GeV$ to 
$O(10^3)~GeV$ or less.}. 
For instance, for $m_{\nu_1} = 0.1~eV$, 
$h_{L,1} = 1$, $M_{H_2} = 300~GeV$, $M_{R1} = 100~MeV$
we find  $  \frac{\delta m^2_{HH_2}}{M^2_{H_2}} = 1.9 \cdot 10^{-8}$ 
and  $\delta m^2_{HH_2} = 1.7 \cdot 10^{-3}~GeV^2 $. 
It is interesting to note that  for $\delta m^2_{HH_2} = 0$
the second Higgs isodoublet vacuum expectation value 
$<H_2> = 0$ at classical level but the spontaneous symmetry breaking of $SU_L(3)\otimes SU_R(3)$ 
chiral symmetry in QCD leads to nonzero vacuum expectation values for the Higgs fields \cite{Tokarev}. 
Really, 
for nonzero  Yukawa interaction $L_{H_2Q_1d} = h_{Qd,11} \bar{Q}_{1L} H_2 d_R + H.c.$    
due to nonzero vacuum expectation value of quark condensate $<\bar{d}d> = - \frac{f^2_{\pi}m^2_{\pi}}{(m_u + m_d)}$ 
($f_{\pi} = 93~MeV$)  
the field $<H_2>$ acquires monzero vacuum expectation value $<H_2 > = \frac{h_{2d_{L}d_{R}}< \bar{d}d >}{ M^2_{H_2}}$.
For example, for  $h_{L,1} = 1$, $M_{H_2} = 300~GeV$ and  $h_{Qd,11} = 10^{-3}$ we 
find\footnote{In our estimate 
 we use the value $m_d + m_u = 11~MeV$}
the  Dirac neutrino mass $m_{\nu_1} \approx 0.1~eV$.  
So for the  model with $\delta m^2_{HH_2} = 0$ the vacuum expectation value $<H_2> = 0$ at 
classical  level but  nonzero quark condensate leads to the appearance of small vacuum expectation value 
$<H_2> \neq 0$ for the second Higgs isodoublet that explains the smallness of the neutrino masses.

It should be noted that the existence of $\Delta S  = 1$ neutral flavour changing interaction (2) leads to 
additional contribution to rare decays $K_L \rightarrow \pi^0 \nu \bar{\nu}$ and 
$K^+ \rightarrow \pi^+ \nu \bar \nu$.   In ref.\cite{GniKr} 
the ratio 
$\beta \equiv \frac{Br^{BSM}(K^+  \rightarrow  \pi^+ \nu_{L1} \bar{\nu}_{R1},  ~\pi^+ \nu_{R1} \bar{\nu}_{L1})                 )
}{Br^{BSM}(K_L  \rightarrow  \nu_{L1} \bar{\nu}_{R1}, ~\nu_{R1} \bar{\nu}_{L1}                               )} $
has been calculated. Here ``BSM'' means the corresponding contribution beyond the SM. 
Note that the  ratio $\beta$ 
does not depend on unknown value of $M_X$ and   on the values of quark masses $m_d, ~m_s$. 
Numerically, for small Majorana neutrino mass 
$\beta \approx 2 \cdot10^{-3}$ \cite{GniKr}. From the experimental 
value of  $ Br(K^+ \rightarrow \pi^+ \nu \bar{\nu})$
  \cite{PartData,PartData1} and its theoretical predictions in the SM \cite{buras} one can deduce 
that  for 
very light $m_{R1} \ll M_{K_L}$ Majorana neutrino  \cite{GniKr} 
\begin{equation}
Br(K_L 
\rightarrow \nu_{L1} \bar{\nu}_{R1}, \nu_{R1} \bar{\nu}_{L1}     ) \lesssim  10^{-7} \,.
\end{equation}
 
For higher  Majorana mass $m_{R1}$   the limit (14) is more weak and 
for the case  $M_{K_L} \geq      m_{{R1}} \geq M_{K^+} -M_{\pi^+}$  when the decay 
$K^+ \rightarrow \pi^+ \nu_{L1} \bar{\nu}_{R1}, \pi^{+}\nu_{R1} \bar{\nu}_{L1} $ is kinematically prohibited but the decay 
$K_L \rightarrow \nu_{L1} \bar{\nu}_{R1}, ~\nu_{R1} \bar{\nu}_{L1}                 $ is still allowed, 
the restriction from $K^+ \rightarrow \pi^+ \nu \bar{\nu}$ decay  does not work.

%  Second model

\section{ $K_L \rightarrow \nu\bar{\nu}$ decay   in 
the $\nu$MSM extension with additional scalar isosinglet}

In this section we discuss the   $K_L$ decay into  neutrinos 
in the $\nu$MSM extension with additional scalar isosinglet field $\phi$.

%The corresponding part of the extended SM Lagrangian is 
%\begin{equation}
%L_{\nu} = \sum_{i,j =1,3}[h_{\nu, i j}\bar{L}_{i}H\nu_{Rj} -M_{i}\frac{\nu_{Ri}\nu_{Ri}}{2} + ~h.c.~] \,.
%\end{equation}
%Here $L_1 = (\nu_{eL},e_L)$, $L_2 = (\nu_{\mu L},\mu_L)$,  $L_2 = (\nu_{\tau L},\tau_L)$ are lepton 
%doublets, $H = (H^0,H^-)$ is the SM Higgs isodoublet and $\nu_{Ri}$ are the Majorana neutrino. As a consequence 
%of see saw mechanism ``active'' neutrino acquire  masses $m_{ij} = m^D_{N}M^{-1}m^D$, where 
%$m^{D}_{ij} = h_{ij}<H>$, $<H> = 174~GeV$. 

The Lagrangian of the model has the form
\begin{equation}
L_{tot} = L_{SM} + L_{Qd\phi} + L_{\nu \phi} + L_{\nu_R} \,.
\end{equation}
Here 
\begin{equation}
L_{Qd\phi} = -\frac{h_{Qd\phi,ij}}{M}\bar{Q}_{Li}\bar{H}\phi d_{Rj} +~H.c.~ \,,
\end{equation}
\begin{equation}
L_{\nu \phi} = \frac{1}{2}\partial^{\mu}\phi\partial_{\mu}\phi -\frac{M^2_{\phi}\phi^2}{2} 
- \lambda\phi^4
-(\frac{\kappa_{i}}{2}\phi \nu_{Ri}\nu_{Ri} + H.c.) \,,
\end{equation} 
\begin{equation}
L_{\nu_R} = i\bar{\nu}_{Rj} \hat{\partial} \nu_{Rj} -(\frac{M_{Rj}}{2}\nu_{Rj}\nu_{Rj} 
 + h_{Lij}\bar{L}_iH\nu_{Rj} + H.c.)  \,,
\end{equation}
where  $L_{SM}$ is the  SM Lagrangian  and   
$L_1 = (\nu_{eL}, e_L)$, $L_2 = (\nu_{\mu L}, \mu_L)$, $L_3 = (\nu_{\tau L}, \tau_L)$,  
  $\bar{H} = (-(H^-)^*, (H^0)^*)$.\footnote{Here $H = (H^0, H^-)$ is the SM Higgs doublet.}

After the spontaneous $SU_L(2) \otimes U(1)$ gauge symmetry breaking the Yukawa interaction  of quarks with 
singlet field $\phi$ is
\begin{equation}
L_{d\phi} = - {\bar{h}_{Qd\phi,ij}}\bar{d}_{Li}\phi d_{Rj} +~H.c.~ \,,
\end{equation}
where 
\begin{equation}
\bar{h}_{Qd\phi,ij} = h_{Qd\phi,ij} \times \frac{<H>}{M} \,
\end{equation}
and $<H> = 174~GeV$.

Note that the interaction (16) is nonrenormalizable. We can consider it as some effective interaction. 
For instance, the effective interaction (16) can be realized in renormalizable extension of the SM model with additional 
scalar field $\phi$ and new massive quark $SU(2)_L$ singlet fields $D_R$, $D_L$ with 
a mass $M_{D}$ and   $U(1)$ hypercharges 
$Y_{D_L} = Y_{D_R} = -\frac{1}{3}$. The interaction of new quark fields  $D_R$, $D_L$ with 
ordinary quarks and the neutral scalar field $\phi$ is
\begin{equation}
L_{qD \phi} = - c_i \bar{Q}_{Li}\bar{H}D_R  -   k_j\bar{D}_Ld_{Rj}\phi + H.c.\,.
\end{equation}

In the heavy $D$-quark mass limit $M_{D} \rightarrow \infty$ we obtain the effective interaction (16) with 
\begin{equation}
\frac{h_{Qd\phi,ij}}{M} = \frac{c_ik_j}{M_D} \,.
\end{equation}

%%% Flavour changing currents

Since  we are interested in the $K_L \rightarrow invisible$ decays we assume that at least one 
Majorana neutrino mass $M_{R1}$ is lighter than  $\frac{M_{K_L}}{2}$. The effective Lagrangian describing the 
decay $K_L  \rightarrow  \nu_{R1} \nu_{R1}$ is
\begin{equation}
L_{ds\nu\nu} = \frac{{\kappa}_{1}}{2M^2_{\phi}}(\bar{h}_{Qd\phi,12}\bar{d}_Ls_R + \bar{h}^*_{Qd\phi,21}\bar{d}_Rs_{L} + H.c.)
(\nu_{R1}\nu_{R1} + \bar{\nu}_{R1} \bar{\nu}_{R1})  \,.
\end{equation}
%Here $M_{\phi}$ is the mass of isosinglet scalar $\phi$.

The  invisible decay $K_L \rightarrow \nu_{R1} \nu_{R1}$ width  is 
determined by formula
\begin{eqnarray}
\Gamma(K_L \rightarrow \nu_{R1}\bar{\nu}_{R1}) =
\frac{M^5_{K_L}}{16\pi M^4_X} \nonumber \\
(\frac{F_K}{2(m_d + m_s)})^2K(M^2_{R_1}/M^2_{K_L}) \,,
\end{eqnarray}
where 
%$K(x) =(1-2x) (1-4x)^{1/2}$ 
$K(x) = (1-4x)^{1/2}$ and 
\begin{equation}
\frac{1}{M^4_X} =  \frac{|{\kappa}^2_{1}(\bar{h}_{Qd\phi,12} 
+\bar{h}_{Qd\phi,21} - \bar{h}_{Qd\phi,12}^*  
-\bar{h}_{Qd\phi,21}^*)^2|}{2M^4_{\phi}} \,.
\end{equation}

% For the $Br(K_L \rightarrow \nu_{R1}\bar{\nu_{R1}}) \geq  10^{-6}$ 
%e can test the value of $M_{X}$ up to
%\begin{equation}
%M_{X} \lesssim  0.6 \cdot 10^5~GeV
%\end{equation}
%for small  Majorana neutrino mass $m_{\nu_{Ri_0}} \ll M_{K_L}$. 

For $\bar{h}_{Qd\Phi,12} = -\bar{h}^{*}_{Qd\Phi,21}$(the matrix $\bar{h}_{Qd,ij}$ is 
antihermitean) the Lagrangian (23) takes the form
\begin{equation}
L_{ds\nu\nu} = \frac{{\kappa}_{1}}{2 M^2_{\phi}}(\bar{h}_{Qd\phi,12}\bar{d}\gamma_5 s  - \bar{h}_{Qd\phi,12}^*\bar{s}\gamma_5 d )
(\nu_{R1}\nu_{R1} + \bar{\nu}_{R1} \bar{\nu}_{R1})  \,
\end{equation}
The Lagrangian (26) does not contain  scalar quark bilinear terms 
$(\bar{d}s + \bar{s}d) (\nu_{R1}\nu_{R1} + \bar{\nu}_{R1} \bar{\nu}_{R1}) $.
As a consequence the decays   
of the $K$ mesons into pions and sterile neutrinos  $K^+ \rightarrow \pi^+ \nu_R\nu_R$,  $K_L \rightarrow \pi^0 \nu_R\nu_R$ 
are absent at least in the leading order on  $|\bar{h}_{Qd\phi,12}|^2$.
  Therefore the  kaon decays 
$K^+ \rightarrow \pi^+ \nu_R\nu_R$,  $K_L \rightarrow \pi^0 \nu_R\nu_R$ don't restrict the 
invisible $K_L \rightarrow \nu_R\nu_R$ decay.

For     $\bar{h}_{Qd\Phi,12} = -\bar{h}_{Qd\Phi,21}^{*}$ the 
exchange of singlet scalar field leads to the 
tree level $\Delta S = 2$ 
interaction  
\begin{equation}
L_{\Delta S = 2} =  \frac{1}{4\Lambda^2_{\Delta S = 2}}\bar{d}\gamma_{5} s \bar{d}\gamma_{5} s   + H.c. \,,
\end{equation}
where
\begin{equation}
 \frac{1}{4\Lambda^2_{\Delta S = 2}} = \frac{\bar{h}^{2}_{Qd\phi,12}}{2 M^2_{\phi}}\,.
 \end{equation}
The  $\Delta S = 2$  bound    (11) restricts rather strongly but not excludes 
the phenomenologically interesting values of the $M_X$. 
For instance, for  $M_{\phi} = 100~GeV$, $M_{R1} = 50~KeV$, $\bar{h}_{Qd\phi,12} = 0.25 i \cdot10^{-5}$ 
 and  $\kappa_{1} = 1$  we find that  $|\Lambda_{\Delta S = 2}| = 2.8 \cdot 10^{7}~GeV $ 
 and  $Br(K_L \rightarrow \nu \nu) \approx 6.5 \cdot 10^{-6}$. 
So we have demonstrated  that the extension of the $\nu$MSM  with aditional isosinglet scalar field 
can lead to the existence of the $K_L \rightarrow invisible$ decay with the 
phenomenologically interesting values of the   $Br(K_L \to invisible)  \geq 10^{-8}$ without 
contradiction with $\Delta S = 2$ bound (11).

In this section  we have considerd the $K_L$ decay into sterile neutrino. It is possible instead 
of sterile neutrino to introduce light dark matter(fermionic or scalar) and 
consider the $K_L$ decay into dark matter particles.  For instance, for $h_{Lij} = 0$ in formula (18) 
we can  consider the $\nu_{R1}$ as stable dark matter particle not directly related with 
left handed neutrinos $\nu_L$. Instead of sterile neutrino $\nu_R$ we can introduce additional light scalar field 
$\chi$. For the interaction
\begin{equation}
L_{\phi\chi\chi} = \lambda_{\phi\chi\chi}\phi\chi^2 \,
\end{equation}    
the invisible $K_L \rightarrow \chi\chi$ decay can occur  if $M_{\chi} < \frac{M_{K_L}}{2}$. 
If the  $\chi$ is stable, it can play the role of light dark  matter.

Note that the $\nu$MSM with additional scalar field can  explain the observed $(g-2)$ anomaly \cite{g-2} if 
we assume the existence of aditional nonzero  interaction of the scalar $\phi$ with charged leptons, 
namely:
\begin{equation}
L_{l\phi H} = -\frac{h_{L\Phi e,ii}}{M_L}\bar{L}_i H \phi e_{Ri} + H.c \,.
\end{equation}  
Here $e_{R1} = e_R$, $e_{R2} = \mu_R$, $e_{R3} = \tau_R$.
After the spontaneous $SU(2)_L\otimes U(1)$ electroweak symmetry breaking the Yukawa interaction of the scalar 
field with charged leptons takes the form
\begin{equation}
L_{l\phi H} = -\bar{h}_{Le,ii}\bar{e}_{Li} \phi e_{Ri} + H.c \,,
\end{equation}  
where 
\begin{equation}
\bar{h}_{Le,ii} = h_{L\Phi e,ii} \frac{<H>}{M_L} \,
\end{equation}
and  $e_{L1} = e_L$, $e_{L2} = \mu_L$, $e_{L3} = \tau_L$.
Note  that in the SM   the renormalizable lepton-Higgs Yukawa interaction 
\begin{equation}
L_{l H} = -{h}_{Lij}\bar{L}_{i}H e_{Rj} + H.c \,,
\end{equation}  
lead to  nonzero   lepton masses due to   nonzero Higgs doublet vacuum expectation 
value $<H> \neq 0$.   Consider the model  with zero $h_{Lij} = 0$ renormalizable Yukawa 
couplings\footnote{We can impose the discrete symmetry $e_{Ri} \rightarrow  -e_{Ri}, 
\Phi \rightarrow - \Phi$ to suppress the renormalizable interaction  (33)}.
For such model  non zero vacuum expectation of the real field $\Phi$ produce  nonzero lepton masses, 
namely 
\begin{equation}
m_{Li} = \bar{h}_{Le,ii}<\Phi>
\end{equation}
Consider the case of real $\bar{h}_{Le,ii}$ coupling constants.   
The additional one loop contribution to muon magnetic moment due to $\phi$ 
scalar exchange is \cite{mureview}
\begin{equation}
\Delta a_{\mu} = \frac{1}{4\pi^2}\frac{m^2_{\mu}}{M^2_{\Phi}} 
\bar{h}^2_{Le,22}[ln(\frac{M_{\Phi}}{m_{\mu}}) 
- \frac{7}{12}],  \, \,
(M_{\Phi} >> m_{\mu}) \,.
\end{equation} 
The precise measurement of the anomalous magnetic
moment of the positive muon from the
Brookhaven AGS experiment \cite{g-2} gives a result which
is $3.6 \sigma$ higher than the Standard Model (SM) prediction, namely 
\begin{equation}
a_{\mu}^{exp} - a_{\mu}^{SM} = (288 \pm 80) \cdot 10^{-11} \,,
\end{equation} 
where $a_{\mu} \equiv \frac{g_{\mu} -2}{2}$. 
  Using the formulae (35, 36) we find that for $m_{\Phi} = 100~GeV$($1~GeV$) the muon 
$g-2$ anomaly can be explained if  $\bar{h}^2_{Le,22} \neq 0$, namely 
\begin{equation}
\bar{h}^2_{Le,22} = (1.6  \pm 0.44) \cdot 10^{-2}\,  ~~for    ~~~~m_{\Phi} = 100~GeV \,,
\end{equation}
\begin{equation}
\bar{h}^2_{Le,22} = (5.9   \pm 1.6) \cdot10^{-6} \, ~~for  ~~~~m_{\Phi} = 1~GeV \,.
\end{equation}

As in the SM the Yukawa couplings $h_{Le.ii}$ are proportional to the lepton masses. 
As a consequence the interaction of the $\Phi $ scalar with electrons is weaker than 
the interaction of the $\Phi $ scalar with muons by factor $m_{\mu}/m_{e} \approx 200$
and the contribution of the $\Phi$ scalar to the electron magnetic moment is 
suppressed at least by factor $(m_{e}/m_{\mu})^2$ in comparison to the muon magnetic moment 
even for superlight $m_{\Phi} \ll m_{e}$ scalar. So the search for light $\Phi$ scalar in 
electron fixed target experiments 
or $e^+e^-$ experiments is very problematic. 
 Light scalar particle $\Phi$ with a mass $m_{\Phi} \lesssim 1~GeV$    decaying into muon pair 
can be searched for at CERN SPS secondary muon beam in full analogy with the search for new 
light vector boson $Z^`$ \cite{GKM}.

\section{Conclusion}
 The observation of the  $K_L \to invisible$ decay  would unambiguously signal the presence of the BSM physics. In this note
we considered   the $K_L \rightarrow \nu\bar{\nu}$ decay in 
the simplest extensions of the SM, such as the 2HDM and the $\nu$MSM with 
additional scalar isosinglet. Using constraints from the 
$\Delta S = 2$ flavour changing interactions and
experimental value for  the  $Br(K^+ \rightarrow \pi^+ \nu \bar{\nu})$ we find that the  $K_L \to invisible$ decay  
branching ratio could be in the   region $Br(K_L \to invisible)\simeq 10^{-8}- 10^{-6}$, 
which is  experimentally accessible,  allowing to test new physics scales well above 100 TeV, which is not accessible at present accelerators. In some scenarios
 the bound from $K^+ \rightarrow \pi^+ \nu\bar{\nu}$ decay can be avoided, as  in the model with the massive  sterile  neutrino. 
This makes  the  $K_L \to invisible$ decay a  powerful clean  probe of  new physics, 
that is complementary to  other  rare $K$ decay channels.  
We  have also demonstrated that the $\nu$MSM 
with additional scalar field can explain the observed muon $(g-2)$ anomaly.
The obtained results  provide a 
strong motivation for a sensitive search for this process in a near future $K$ 
decay experiment proposed in \cite{Gninenko}. It should be noted that in 
full analogy with the case of $K_L$ invisible decay we can  expect the existence of 
invisible decays of $B_d$ and $B_s$ mesons, see e.g. \cite{alex,babar}, 
 with the branchings similar to those discussed above.

\end{document}